\documentclass{aa}
\usepackage{graphicx}
\begin{document}

\title{Meissner Effect and Vortex Dynamics in Quark Stars}
\subtitle{A Model for Soft Gamma-Ray Repeaters}

\author{R. Ouyed~\inst{1}
\and {\O}. Elgar{\o}y~\inst{2,3}
\and H. Dahle~\inst{3}
\and P. Ker{\"a}nen~\inst{2}
}

\institute{Department of
Physics and Astronomy, University of Calgary, 2500 University Drive NW, Calgary, Alberta, T2N 1N4 Canada 
\and Nordic Institute for Theoretical Physics, Blegdamsvej 17,
DK-2100 Copenhagen, Denmark \and
Institute of Theoretical Astrophysics, University of Oslo, P.O. Box 1029, Blindern, N-0315 Oslo, Norway 
}

\offprints{ouyed@phas.ucalgary.ca}

\date{Received/Accepted}

\abstract{We present a new model for soft gamma-ray repeaters based on a quark star born with 
temperatures above the critical value ($T_c$) for the onset of the colour-flavor locked 
superconductivity. The quark star then quickly cools below $T_c$, expelling a fraction of 
the surface magnetic field via the Meissner effect. We show that if a small fraction 
($\leq 10\%$) of the surface magnetic field ($10^{14} - 10^{15}\, {\rm G}$) is expelled, it quickly decays via
magnetic reconnection and heats up the quark star surface to temperatures $> 10^9\,{\rm K}$. 
Created $(e^{+},e^{-})$ pairs annihilate into gamma rays emitted in a giant burst (the first 
burst in our model), with a luminosity of $\sim 10^{45}\, {\rm ergs}\, {\rm s}^{-1}$. 
Subsequent bursts result from the restructuring of the surface magnetic field
 following the formation and relaxation of a vortex lattice 
which confines the internal magnetic field.  During
this phase, energy is sporadically released as a consequence of 
magnetic reconnection events in the entangled surface magnetic field as it evolves into a smooth, more stable, configuration. 
The star eventually enters a quiescent phase in which energy is continuously supplied by vortex annihilation at the surface.   
As the star spins down, the outermost vortex lines will be pushed to the surface where they annihilate and release their 
confined magnetic field. We show that the corresponding luminosity is $L_v \sim 10^{36}\, {\rm ergs}\, {\rm s}^{-1}$ for 
a typical soft gamma-ray repeater spinning with a period of $8\, {\rm s}$ and a surface magnetic field not exceeding $10^{15}\, {\rm G}$.
Our model can be applied to any situation where a $T>T_{\rm c}$ quark star is
 generated.
We discuss the connection between anomalous X-ray pulsars and soft gamma-ray repeaters in the context of our model. 
}

\authorrunning{Ouyed et al.}

\maketitle

\section{Introduction}

Soft $\gamma$-ray repeaters (SGRs) are sources of recurrent, short ($t \sim 0.1$s), intense
($L \sim 10^3 - 10^4 L_{\rm Edd}$) bursts of $\gamma$-ray
emission with a soft energy spectrum.
The normal pattern of SGR activity are intense activity periods which can last weeks or
months, separated by quiescent phases lasting years or decades. The five known SGRs are located in our
Galaxy, or, in the case of SGR 0526-66, in the Large Magellanic Cloud. 
The two most intense SGR bursts ever recorded were the 5 March 1979 giant
flare of SGR 0526-66 (Mazets et al.\ 1979) and the similar 27 August 1998 giant flare of SGR 1900+14. 
The peak luminosities of these events ($~\sim 10^6 -10^7 L_{\rm Edd}$)
exceeded the peak luminosities of ``normal'' SGR bursts by a factor $>10^3$.
In Table~\ref{tab:burstprop} we summarize the burst properties of these
giant flares.

\begin{table}
{\parindent=-0.2in
\caption{Giant burst properties.}
\begin{tabular}{ccc}
Object & SGR 0526-66 & SGR 1900+14 \\
\hline
Active periods & 1979-83 & 1979, 1992, 1998-99 \\
Giant Burst & 5 March 1979 & 27 August 1998 \\
\hline
\hline
Precursor: &  &  \\
Duration, s & no data  & $\sim 0.05$ \\
$kT$ (keV) &  & $\sim 20$ \\
\hline
Hard $\gamma$-ray spike: &  & \\
Duration, s & $\sim 0.25$ & $\sim 0.35$ \\
Peak luminosity, ergs s$^{-1}$ & $1.6 \times 10^{45}$ & $\ga 3.7 \times
10^{44}$ \\
Energy release, ergs & $1.3 \times 10^{44}$ & $\ga 6.8 \times 10^{43}$ \\
$kT$ (keV) & $\sim 246$ & $\sim 240$ \\
\hline
Bright X-ray emission: &  &  \\
Duration, s & $\sim 180$ & $\sim 370$ \\
Pulsation period, s & $8.1$ & 5.16 \\
Energy release, ergs & $3.6 \times 10^{44}$ & $ 5.2 \times 10^{43}$ \\
$kT$ (keV) & $\sim 30$ & evolves, $31.2 \rightarrow 28.9$  \\
\hline
\end{tabular}
}
\note{Data sources: Cline et al. 1980; Fenimore, Klebesadel, \& Laros
1996; Feroci et al. 2001; Hurley et al. 1999a; Ibrahim et al.\ 2001;
Mazets et al. 1999. ``No data'' means that Mazets et al.\ 1979 have not
plotted any data for the last second before the burst (the time resolution
for their observations of the pre-burst background is apparently $> 1$s).}
\label{tab:burstprop}
\end{table}

Several SGRs have been
found to be X-ray pulsars with an unusually high spin down rate of
$\dot{P} / P \sim 10^{-10}$~s$^{-1}$, usually attributed to magnetic braking caused by a super-strong
magnetic field $B > 10^{14}$G, which implies that SGRs are magnetars (Golenetskij et al.\ 1979;
Duncan \& Thompson 1992, Kouveliotou et al.\ 1998, Kouveliotou et al.\ 1999). 
In the magnetar model, the magnetic field is
the likely provider of the
burst energy, since it is the dominant source of free energy in the star.
A common scenario is that stresses build up in the magnetic field and create
a quake in the crust of the neutron star which ejects hot plasma Alfv{\'e}n waves through
its rigid magnetosphere (Thompson \& Duncan 1995; 1996). The magnetic field of such a star would
have grown to magnetar-scale strengths because of strong convection during the collapse of the
proto-neutron star core (Duncan \& Thompson 1992; Thompson \& Duncan 1993).

In this paper, we propose an alternative model where the SGR activity is produced by phenomena
occurring in a hot ($T>T_{\rm c}$) quark star.
In \S~\ref{sec:cooltime} we 
discuss the cooling timescales of the quark star and 
 the onset of colour superconductivity. 
In \S~\ref{sec:expulsion} we show how giant flares such as the 5 March 1979
event could be powered by the expulsion of a fraction of the magnetic field due to the Meissner effect
after the onset of superconductivity. We also calculate the light curve and temperature evolution for
such a burst in our model. The remaining magnetic field in the star is confined to vortices, which
will evolve into a lattice configuration, as discussed in \S~\ref{sec:vortices}. Subsequent weaker SGR
bursts may be due to the reorganization of the external magnetic field, and
the luminosity of SGRs in their quiescent phase may be mostly powered by the annihilation
of vortices at the stellar surface, as detailed in \S~\ref{sec:quiescent}.
After decades of quiescence, a crust may form, leading to occasional crustquakes which could power
later burst events, and additional events could be due to rare impacts; such events are discussed in
\S~\ref{sec:laterbursts}. In \S~\ref{sec:SGRdisc}, we briefly discuss
known SGRs within our model, and in \S~\ref{sec:discussion}  
we discuss the suggested connection SGRs have with anomalous X-ray pulsars (AXPs) in relation to our model.   
Finally, we conclude in \S~\ref{sec:conclusion}.

\section{Cooling timescale and onset of colour superconductivity: {\it the CFL star}}
\label{sec:cooltime}

Present estimates of the critical temperature for onset of 
colour-flavor locked (CFL) superconductivity suggest $T_{\rm c} \sim 10^{12}\;{\rm K}$ 
(Rajagopal \& Wilczek 2000)
whereas early estimates indicated $T_{\rm c} \sim 10^{9}\;{\rm K}$ 
(Bailin \& Love 1984). The quark star we
take to be born  with an effective temperature higher than $T_{\rm c}$.
Once formed, the quark star will cool rapidly through neutrino
emission.     
Using the result of Haensel (1991) for the neutrino emissivity, 
\begin{equation}
\varepsilon_\nu = 2.2 \times 10^{26} \alpha_c Y_e ^{1/3}(n_b/n_0)
T_9^6 \;{\rm ergs}\,{\rm cm}^{-3}\,{\rm s}^{-1},  
\label{eq:heat4}
\end{equation}
where $\alpha_c = g^2/4\pi$ is the QCD fine structure constant, $g$ is
the quark-gluon coupling constant, $Y_e = n_e / n_b$ is the  ratio 
of electron and baryon numbers, $n_0 \simeq 1.7\times 10^{38}\;{\rm cm}^{-3}$ is the nuclear 
matter saturation density, $T_9$ is the temperature in units of
$10^9\;{\rm K}$, and taking the specific heat per volume as 
(Iwamoto 1982) 
\begin{equation}
C_q = 2.5 \times 10^{20}(n_b/n_0)^{2/3}T_9\;{\rm ergs}\,{\rm cm}^{-3}
\,{\rm K}^{-1}, 
\label{eq:heat2}
\end{equation}
we can estimate the timescale for neutrino cooling from an initial
temperature $T_{9,\rm i}$ to a final temperature $T_{9,\rm f} \ll T_{9,\rm
  i}$:  
\begin{equation}
\tau_{\rm cool} \simeq 3\times 10^2\alpha_c^{-1}Y_e^{-1/3}
\left(\frac{n_b}{n_0}\right)^{-1/3}T_{9,\rm f}^{-4}\;{\rm s},
\label{eq:cooltime}
\end{equation}
According to this estimate a critical temperature of, say, $T_c \sim
10^{11}\;K$  will be reached within $\tau_{\rm cool} \sim 0.4\;{\rm
  ms}$.  Thus, the star will 
rapidly undergo  a phase transition to a colour superconducting state 
(except in the cores of the
vortices formed in response to the rotation of the star; see \S~\ref{sec:vortices}). 

\section{Giant Burst: {\it Expulsion and decay of the magnetic field}} 
\label{sec:expulsion}

In an ordinary superconductor there is a thermodynamical critical 
field $H_{\rm c}$ (or $H_{c1}$ for type II superconductors),  
determined by the free-energy difference between the normal and 
superconducting states in zero field, below which a magnetic field 
will be screened from the interior of the superconductor: this is 
the Meissner effect.  The existence of the magnetic 
Meissner effect in colour superconductors is a matter of some debate 
(Alford, Berges \& Rajagopal 2000, Sedrakian et al. 2001, Iida \& Baym
2002) but we will proceed on the assumption 
that there is a Meissner effect in the CFL phase, and that at least 
part of the magnetic field is expelled from the surface layers of 
the star.   

The model we consider assumes that the star 
intially consists of uniform density strange quark matter with an associated 
surface electric field \cite{alcock86}. 
In this situation the calculations of \cite{pu2002} show that the
surface of the star cools faster than the core because of thermal
emission of photons and $e^+\, e^-$-pairs from the surface. Therefore
the CFL phase transition occurs first in a thin layer of thickness 
$\delta \ll R$ ($R$ is the radius of the star) close to the surface.
The magnetic field decay discussed later will thus heat the surface
leading to thermal emission of photons.

The penetration depth $\lambda$ for the magnetic field 
is of the order 1 fm (see Iida and Baym 2002) so the 
magnetic field is negligible throughout the superconducting layer provided that 
$\delta \gg \lambda \sim \;{\rm fm}$.  Note that the thermodynamical critical 
field is $H_{\rm c}\sim 10^{19}\;{\rm G}$ (Iida \& Baym 2002), 
much larger than the field strengths of interest to us.  

\begin{figure}[b!]
\resizebox{\hsize}{!}{\includegraphics[height=1cm, width=1cm]{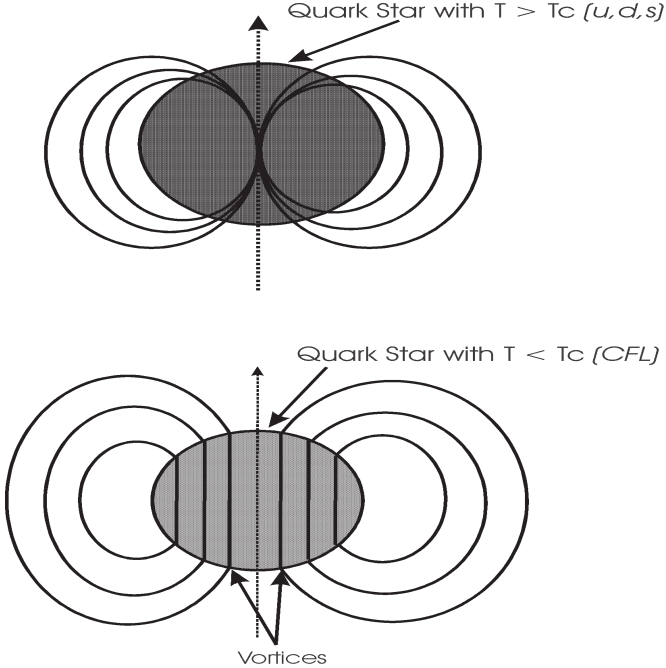}}
\caption[Meissner]
{Schematic view of the initial and final stages in the development of the SGR
in our model. The QS, born with $T > T_c$  quickly cools into a CFL star. In the 
early stages of this transition, a fraction of the surface magnetic field is 
expelled by the Meissner effect, inducing magnetic
reconnection events  leading to the main burst. 
In the late stage, once the vortices form (parallel to the rotation axis), 
the system evolves into a configuration  where most of the magnetic field is 
confined to the vortex lines. Magnetic reconnection events between the complex 
surface magnetic field and the external part of the magnetic field coupled to 
the vortex lines lead to occasional flares (subsequent bursts).
In the SGR quiescent phase, magnetic energy is released as vortex lines are 
being continuously pushed to the surface and annihilate there. 
In the more general case where the magnetic field is not aligned with 
the rotation axis, we expect an even  more complex behaviour/evolution 
of the field leading to more reconnection events.} 
\label{fig:Meissner}
\end{figure}

\subsection{Expulsion timescale}

The time scale for the expulsion is determined by the conductivity 
of the normal, non-superconducting state according to 
\begin{equation}
\tau_{\rm exp} = \frac{4\pi \sigma_{\rm el} d^2}{\pi c^2}\left(\frac{B}{H_c}\right)^2,
\label{eq:texp}
\end{equation} 
where $\sigma_{\rm el}$ is the electrical conductivity in the normal 
state while $B$, $c$ and $d$ are the magnetic field strength,
the speed of light and the thickness of the
superconducting layer, respectively.   For quarks, this quantity was    
found by Heiselberg \& Pethick (1993) to be given by 
\begin{equation}
\sigma_{\rm el,quark} \simeq (\alpha_s T_9)^{-5/3}\left(\frac{\mu}{300\;{\rm
    MeV}}\right)^{8/3} \times 10^{19}\;{\rm s}^{-1}, 
\label{eq:elcond}
\end{equation}
where $\mu$ is the quark chemical potential.  
For $\alpha_s=0.1$ (strong coupling constant), $\sigma_{\rm el}$ varies from $\sim 10^{17}\;{\rm
  s}^{-1}$ for $T=10^{11}\;{\rm K}$ to $\sim 10^{14}\;{\rm s}^{-1}$
  for $T=10^{13}\;{\rm K}$.
The electron contribution to the electrical conductivity can be 
estimated from the classical result 
\begin{equation}
\sigma_{\rm el,electron} \simeq \frac{8\pi\alpha n_e
  \ell_e}{m_e\overline{v}_e},
\label{eq:texp2}
\end{equation}
where $\alpha\simeq 1/137$ is the QED fine structure constant, $n_e$ 
is the electron number density ($m_e$ is the electron mass), $\ell_e$ is the electron mean free
path, $\overline{v}_e$ their mean thermal velocity, and taking  
the ultrarelativistic limit where $\overline{v}_e \sim c$ and $m_e$ 
is replaced by $T$, using $\ell_e \sim n_e^{-1/3}$ as a crude estimate, 
this gives 
\begin{equation}
\sigma_{\rm el} \simeq 7.1\times 10^{21}
\left(\frac{Y_e}{10^{-4}}\right)^{2/3}\left(\frac{n_b}{n_0}\right)^{2/3}
\left(\frac{1\;{\rm MeV}}{T}\right)\;{\rm s}^{-1}.
\label{eq:texp3}
\end{equation} 
More refined estimates (Baym \& Heiselberg 1997; 
Arnold, Moore \& Yaffe 2000; Shovkovy \& Ellis 2003) give 
$\sigma_{\rm el}\sim 10^{23}$--$10^{24}\;{\rm s}^{-1}$.   
 With $H_c = 10^{19}\;{\rm G}$ we obtain 
\begin{equation}
\tau_{\rm exp} \simeq 10^{-6}\left(\frac{\sigma_{\rm
      el}}{10^{17}\;{\rm s}^{-1}}\right)\left(\frac{\delta}{10^3\;{\rm
      cm}}\right)^2\left(\frac{B}{10^{15}\;{\rm G}}\right)^2\;{\rm s}, 
\label{eq:texp4}
\end{equation}
so a reasonable estimate is $\tau_{\rm exp}$ of the order of a few 
seconds or less  since $\delta_{\rm max.}\sim R\sim 10$~km.

\subsection{Energetics}

The expelled magnetic field
would quickly lead to magnetic reconnection in the magnetosphere.
 Since reconnection typically occurs
at a fraction of the Alfv\'en velocity, the growth
time of the instability can be estimated
to be $\sim 10^{-5}-10^{-4}$ s which is comparable
to the $\sim 2\times 10^{-4}$ s rise time
of the March 5 event (Mazets et al. 1979;
Paczy\'nski 1992). The long expulsion time (see Eq. \ref{eq:texp4} 
for $\delta_{max.}\sim R$)
implies that the reconnection events continued
for many times the Alfv\'en time scale which
we suggest could account for 
the longer duration of the hard transient phase of the burst.

The energy released is 
\begin{equation}
E_{\rm M} \sim 1.7 \times 10^{47}\, {\rm ergs} \left(\frac{\beta B}{10^{15}\, {\rm G}}\right)^{2}\left(\frac{R_{\rm QS}}{10\, {\rm km}}\right)^{3}\ ,
\end{equation}
where $\beta$ is the fraction of the surface magnetic
field expelled with the subsequent reconnection events (the subscript
``M" stands for main). 
Locally we expect the energy from the magnetic reconnection event to be 
rapidly converted to a thermal photon-pair plasma (e.g. Thompson and 
Duncan 1995, section 3) which heats up the surface.  
The cooling is defined by the rate at which 
the photons escape this heated region.  However, since these 
reconnection events occur very close to the surface of the star, 
we expect the photons to be trapped, and the cooling time 
to be long enough to allow for thermalization.  
The corresponding thermal temperature is 
\begin{equation}
T_{\rm M} \sim 4.0\times 10^{9}\, {\rm K} \left(\frac{\beta
  B}{10^{15}\, {\rm G}}\right)^{1/2} \left(\frac{R_{\rm QS}}{10\, {\rm
    km}}\right)^{1/4}\ , 
\label{eq:tsurf}
\end{equation}
which is the maximum surface temperature that can be reached.  
Even smaller surface temperatures (when cooling is faster  
than heating) are enough to trigger the mechanism of thermal emission, 
as we discuss next.  

\begin{figure}[]
\resizebox{\hsize}{!}{\includegraphics{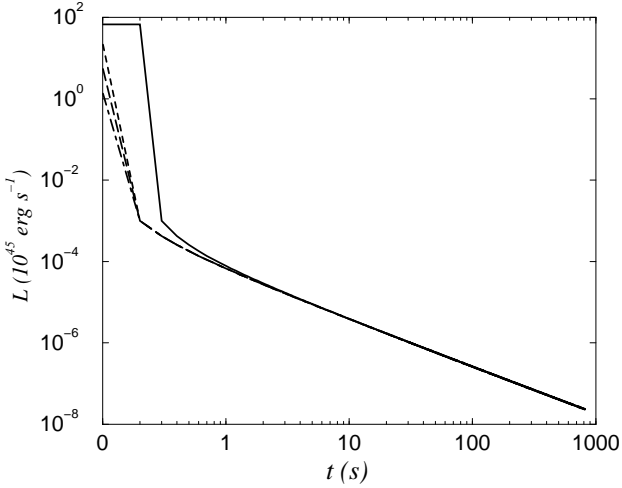}}
\caption[Light curve]
{The light curve expected in our model. The four curves (from top to
  bottom) correspond to $\beta = 0.5$, $0.2$,  $0.1,$ and $0.05$.}
\label{fig:lightcurve}
\end{figure}

\subsection{Light curve}

Usov (2001) showed that creation of $e^+e^-$ pairs by the Coulomb
barrier at the quark star surface is the main mechanism of thermal
emission from their surface at the temperature $T_{\rm S} < 5\times 10^{10}$ K. 
Created $e^+e^-$ pairs mostly annihilate in the vicinity of the quark
star into $\gamma$-rays, and Usov (2001) argued that the light curves of
the March 5 1979 and August 27 1998 events may be explained in a model
where the burst radiation is produced by the bare surfaces of such
stars heated up to $\sim 2\times 10^{9}$ K by impacts of massive
cometlike objects.  He also points out that any other mechanism 
which quickly heats up the surface can explain the events, so  
Eq. (\ref{eq:tsurf}) indicates that the Meissner effect and 
subsequent decay of the expelled magnetic field can power a burst. 
To give further evidence for this we have computed the light curves 
in our model.  

We consider a thin layer of thickness $\delta \ll R$ 
close to the surface of the star which is heated by the 
release of magnetic energy $E_{\rm dec}$ from the decaying 
magnetic field.  The temperature in the layer (taken to be 
isothermal) is governed by the equation 
\begin{equation} 
V_\delta C \frac{dT}{dt} = Q - V_{\delta}\tilde{\epsilon}_\nu, 
\label{eq:heat5}
\end{equation}
where $V_\delta = 4\pi R^2 \delta$ is the volume of the layer, and 
$Q$ for times $0 < t < \tau_{\rm dec}$ is given by the energy released 
per unit time by the decay of the magnetic field, $Q \simeq E_{\rm
  dec} / \tau_{\rm dec}$, while for times $ t > \tau_{\rm dec}$ 
it is given by the luminosity of the $e^+ e^-$ pair emission, 
$Q = - 4\pi R^2 \epsilon_\pm f_\pm$, where $\epsilon_\pm = m_e c^2 
+ kT $ is the mean energy of the electron-positron pairs created, 
\begin{equation}
f_\pm = 1.6 \times 10^{39}T_9^3 \exp\left(-\frac{11.9}{T_9}\right)
J(\zeta)\;{\rm cm}^{-2}\,{\rm s}^{-1}\ , 
\label{eq:heat6}
\end{equation}
is the flux of pairs per unit surface area, 
\begin{equation}
J(\zeta) =
\frac{1}{3}\frac{\zeta^3\ln(1+2\zeta^{-1})}{(1+0.074\zeta)^3}
+\frac{\pi ^5}{6}\frac{\zeta^4}{(13.9+\zeta)^4}, 
\label{eq:heat7}
\end{equation}
and $\zeta = 20 / T_9$.  The specific heat per unit volume $C$ is 
the sum of the contribution from the electrons 
(Blaschke, Grigorian, \& Voskresensky 2001)  
\begin{equation}
C_e \simeq 5.7\times 10^{19}Y_e^{2/3}(n_b/n_0)^{2/3}T_9\;{\rm ergs} 
\,{\rm cm}^{-3}\,{\rm K}^{-1},
\label{eq:heat8}
\end{equation}
and the heat capacity of the quarks in the CFL phase is modified from
Eq.(\ref{eq:heat2}) according to 
\begin{equation}
\tilde{C}_q = C_qf(T/T_c),
\label{eq:heat9}
\end{equation}
where $C_q$ is given by Eq. (\ref{eq:heat2}) and 
\begin{equation}
f(x) = \frac{3.2}{x}\exp\left(\frac{1}{0.57x}\right)(2.5-1.5x+3.6x^2),
\label{eq:heat10}
\end{equation} 
see Horvath, Benvenuto, \& Vucetich (1991).  Also, the neutrino 
luminosity (\ref{eq:heat4}) is suppressed in the CFL phase by the 
Boltzmann factor $\exp(-\Delta/k_{\rm B}T)$, where $\Delta$ is the 
energy gap, i.e. 
\begin{equation}
\tilde{\epsilon}_\nu = \epsilon_\nu g(T/T_c),
\label{eq:heat11}
\end{equation}
where 
\begin{equation}
g(x) = \exp\left(-\frac{\sqrt{1-x}}{0.57x}\right),
\label{eq:heat12}
\end{equation}
and we have assumed that the temperature dependence of the 
energy gap is $\Delta(T) = \Delta(T=0)(1-T/T_c)^{1/2}$ (e.g. 
Carter \& Reddy 2000), and used the standard relation 
$k_{\rm B}T_c \simeq 0.57\Delta(T=0)$.  

\begin{figure}[t!]
\resizebox{\hsize}{!}{\includegraphics[width=4cm,height=3cm]{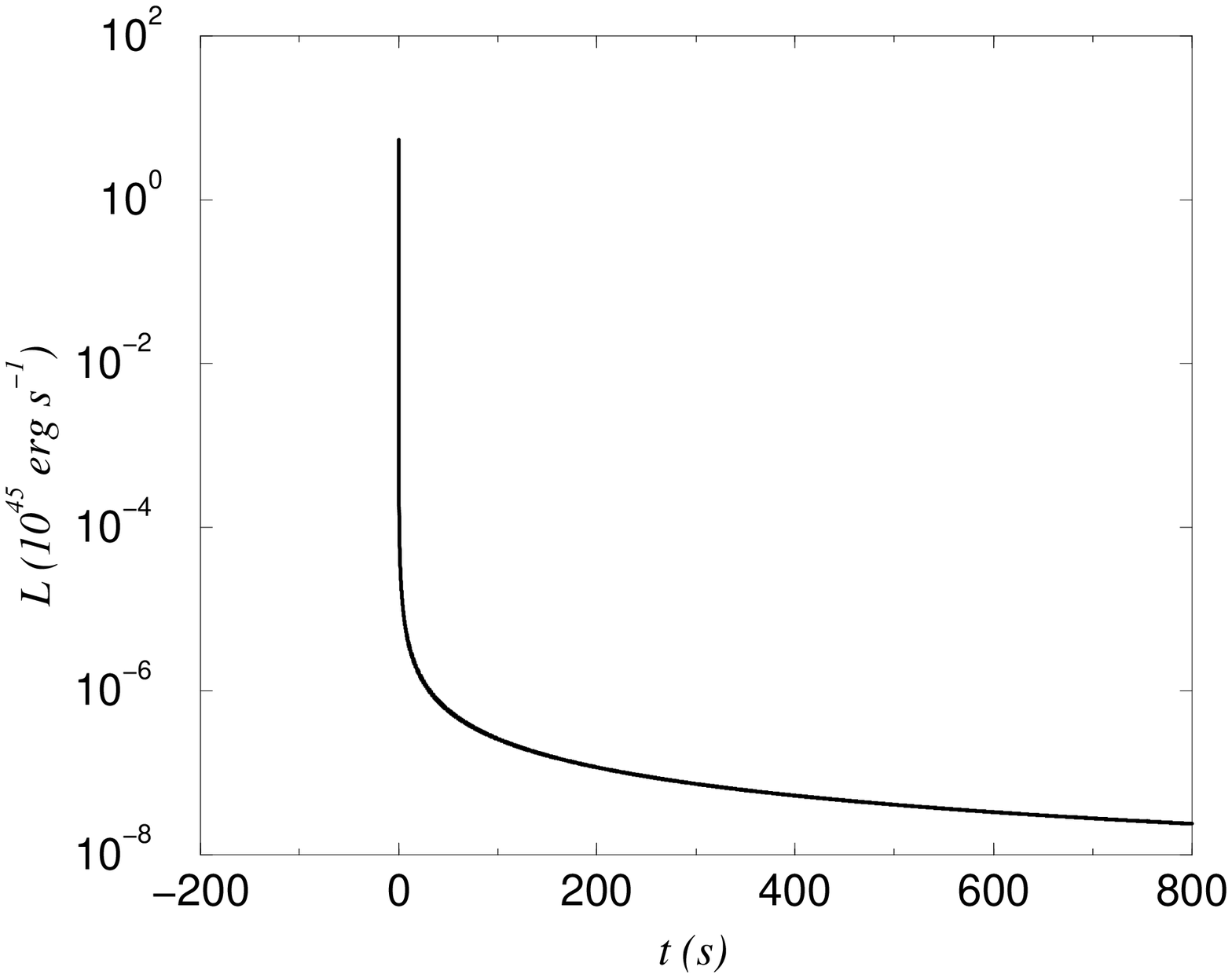}}
\resizebox{\hsize}{!}{\includegraphics[width=4cm,height=3cm]{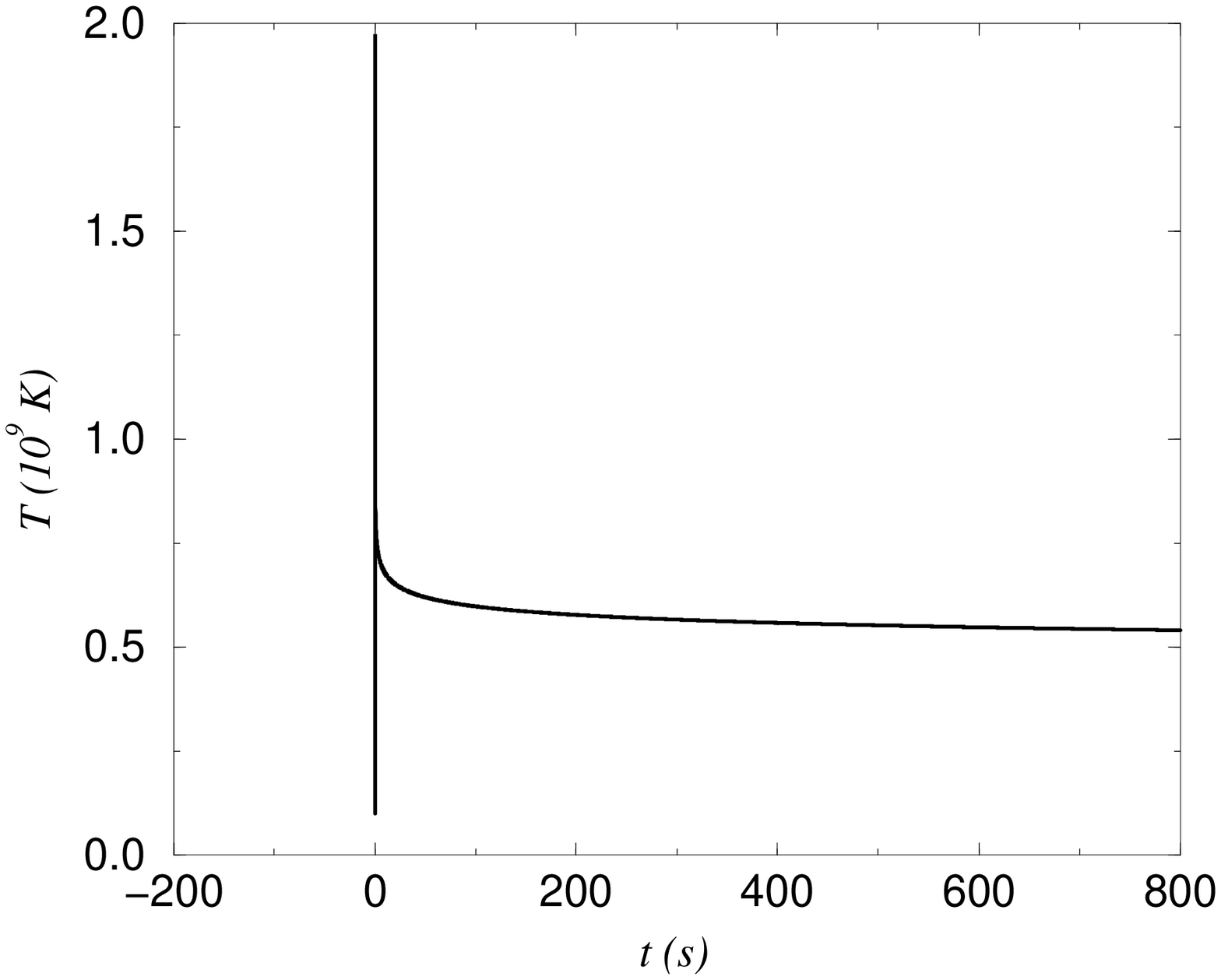}}
\caption[Light curve]
{Upper panel: The light curve for $\beta = 0.1$. Lower panel: The corresponding time evolution of the surface temperature.}
\label{fig:lightcurve_0.1}
\end{figure}

We have solved equation (\ref{eq:heat5}) numerically 
with the initial condition $T = 10^8\;{\rm K}$.  
For the parameters we used the values $\alpha_c=0.1$, $n_b=5n_0$, 
$Y_e=10^{-4}$, $R=10^6\;{\rm cm}$, $\delta = 10^3\;{\rm cm}$, 
$B = 10^{15}\;{\rm G}$, $\tau_{\rm dec}=0.1$.  
We varied the fraction of the magnetic 
field expelled by the partial Meissner effect, using $\beta=0.5$, 
$0.2$, $0.1$ and $0.05$ as representative values.   
The light curves are shown in Figure \ref{fig:lightcurve}.  
The maximum luminosity for $\beta=0.1$ is $5\times 10^{45} \;{\rm
  ergs}\,{\rm s}^{-1}$, so the decay of 10 \% of a surface magnetic
field of $10^{15}\;{\rm G}$ is enough to power the main burst.  
The light curve and time evolution of the surface temperature for this 
case are shown in Figure \ref{fig:lightcurve_0.1}.

We note that since the CFL phase transition occurs only once, 
there will be only one giant burst in our model.
We also  note that our model is too simplistic to reproduce
the periodic pattern (e.g. the 8.0 s period in the 
March 5 event) overimposed to a smooth exponential decay. However,
given the partial Meissner effect and the resulting
random reconnection fronts we expect isolated spots 
on the surface of the star to be heated and release energy
as described above. These hot spots 
would pulse at the same rate as the rotation of the star, and thus,
would appear periodic. We further note that a few of these
hot spots could appear simultaneously at random locations on the 
surface of the star. This means that each of the pulses (superimposed
on the smooth curve) could consist of even smaller subpulses.
Multiple small hot spots should produce many harmonics 
in the lightcurve.  
In other words, if our model is a correct representation of SGRs, 
observations could constrain the number of hot spots. 

Finally, we note that given the rapid cooling time scale of a quark 
star, as discussed in section 2, the CFL phase transition should happen 
immediately after the formation of the quark star.  However, this 
does not necessarily mean that the giant burst will follow 
immediately after the core collapse in a supernova since the quark star 
can in principle be formed also much later.

\section{Subsequent bursts: {\it Vortex lattice dynamics}}
\label{sec:vortices}

\subsection{Vortex formation and relaxation}

The CFL front quickly expands to the entire star followed
by the formation of rotationally induced vortices,
analogously to rotating superfluid He$^3$ (the vortex lines are parallel to the rotation axis; Tilley\&Tilley 1990). 
Via the Meissner effect, 
the magnetic field is partially screened from the regions outside the vortex cores. The system now consists of alternating 
regions of superconducting material with a screened magnetic field and the vortices where most of the magnetic field resides. 

The vortex relaxation time, which is within an order of magnitude similar to the random diffusion timescale, can be estimated to be
\begin{equation}
\tau_r \simeq \frac{R^{2}}{\lambda_{\rm v} c}\ , 
\end{equation}
\noindent
where $\lambda_{\rm v} = (\pi R^{2} / N_{\rm v})^{1/2}$ is the vortex mean free path, and $N_{\rm v}$ is the number of vortices in the entire star,
\begin{eqnarray}
N_{\rm v} &=& \frac{2}{3} \mu R^2 \Omega  \nonumber \\
& \simeq & 6.4 \times 10^{15} \left(\frac{1 {\rm s}}{P}\right) \left(\frac{\mu / 3}{300 {\rm MeV}}\right) \left(\frac{R}{10 {\rm km}}\right)^2\ ,   
\end{eqnarray}
\noindent
(Iida \& Baym 2002), where $\mu$ is the quark chemical potential related to the density. 
Using the above values, we thus estimate a relaxation timescale of 
$\sim 11\;{\rm min}$ for $P = 5\;{\rm s}$.  Note that this is within 
an order of magnitude consistent with the relation between lattice 
formation time and rotation period found in numerical simulations of 
vortices in rotating Bose-Einstein condensates (Tsubota, Kasamatsu, \&
Ueda 2002).  

\subsection{Magnetic field restructuration and reorganization}

Given the complicated structure of the resulting surface magnetic field,
following relaxation, 
the latter may then suffer from frequent magnetic reconnections and thus account for the
subsequent bursts. This phase could last for days or months, depending on the  
magnetic field strength, the period and the fraction of the field which remained entangled in the period immediately following 
the giant burst. These random reconnection events
would bear many (temporal and spectral) similarities to the 
main burst but we expect them
to be less energetic as the magnetic field slowly decays and weakens. 
Eventually, the magnetic field evolves into a stable configuration (see Figure~\ref{fig:Meissner}) after which the star enters
a quiescent phase.

\section{Quiescent phase: {\it Vortex annihilation}} 
\label{sec:quiescent}

The number of vortices decreases with the spin-down of the star, and can be formulated as (extrapolating from results established for
cylindrically rotating superfluids):\footnote{In
rotating superfluids the density of vortex lines in equilibrium rotation is $n_{\rm v} = \frac{2 \Omega}{\kappa}$, where $\kappa = \nu h / 2 m$ is the circulation of a vortex with quantization number $\nu$ for
atoms of mass $m$ (Ruutu et al.\ 1997);
$h$ is the planck's constant. In the continuum limit a totally filled cylindrical container with radius R would have $N_0 = \pi R^2 n_{\rm v}$ lines. Interactions with the lateral walls give rise
to an annular vortex-free region along the wall. Its width $\lambda$ is of order the intervortex distance $r_{\rm v} = (\kappa / 2 \pi \Omega )^{1/2}$ (expressed here as the radius of the Wigner-Seitz unit cell of the vortex lattice). As a result, the total line number $N (\Omega ) \simeq N_0 (1 - 2 \lambda_{\rm v} / R)$ is always less than $N_0 ( \Omega )$.}  
\begin{equation}
\frac{dN_{\rm v}}{d\Omega} = \frac{N_0}{\Omega} \left[ 1 - \frac{\lambda_{\rm v}}{R} \right] \simeq \frac{N_0}{\Omega}\ . 
\end{equation}
As the star spins down, the outermost vortex line will be pushed to the surface and annihilate there (Ruutu et al.\ 1997). 
The corresponding luminosity is 
\begin{equation}
L_{\rm v} = \frac{dE_{\rm v}}{dt} = \epsilon_{\rm v} \frac{dN_{\rm v}}{dt} = \epsilon \frac{dN_{\rm v}}{d\Omega} \frac{d \Omega}{dt} = \epsilon_{\rm v} N_0 \frac{\dot{\Omega}}{\Omega}\ ,
\label{eq:luminosity}
\end{equation}
where $\epsilon_{\rm v} = \epsilon_0 + \epsilon_{\rm m} \simeq \epsilon_{\rm m}$, 
is the energy density per vortex which consists of the
rest mass energy ($\epsilon_0$) and the
confined magnetic energy ($\epsilon_{\rm m}$). This
means $\epsilon_{\rm v} N_0 \simeq E_{\rm mag} \simeq 1.7 \times 10^{47} {\rm ergs} \left(B/10^{15}{\rm G}\right)^2 \left(R/10 {\rm km}\right)^3$ (recall that most of the magnetic field/energy resides in the vortex and that $\epsilon_{\rm m}>>\epsilon_{\rm v}$). The star continues to loose rotational energy due to the electromagnetic radiation losses which allow us to write in the simplest
approximation, 
\begin{eqnarray}
\frac{\dot{\Omega}}{\Omega} & = & - \frac{4 B^2 R^6 \Omega^2}{9 I c^3} \simeq 3.24 \times 10^{-10} {\rm s}^{-1} \times \nonumber \\ 
& \times & \left(\frac{B}{10^{15} {\rm G}}\right)^2 \left(\frac{R}{10 {\rm km}}\right)^4 \left(\frac{1 {\rm s}}{P}\right)^2 \left(\frac{M_{\sun}}{M}\right)\ , 
\label{eq:spindown}
\end{eqnarray}
where $I$ is the star's moment of inertia (Manchester \& Taylor 1977). 

Combining equation~(\ref{eq:luminosity}) and equation~(\ref{eq:spindown}), the luminosity induced by vortex annihilation is 
\begin{eqnarray}
L_{\rm v} & = & 5.51 \times 10^{37} {\rm ergs}\, {\rm s}^{-1} \times \nonumber \\ 
& \times & \left(\frac{B}{10^{15}{\rm G}}\right)^4 \left(\frac{R}{10 {\rm km}}\right)^5 \left(\frac{1 {\rm s}}{P}\right)^2 \left(\frac{M_{\sun}}{M}\right)\ . 
\label{eq:qluminosity}
\end{eqnarray}
\noindent
This can be compared to the luminosity from the dipole radiation, 
\begin{figure}
\resizebox{\hsize}{!}{\includegraphics{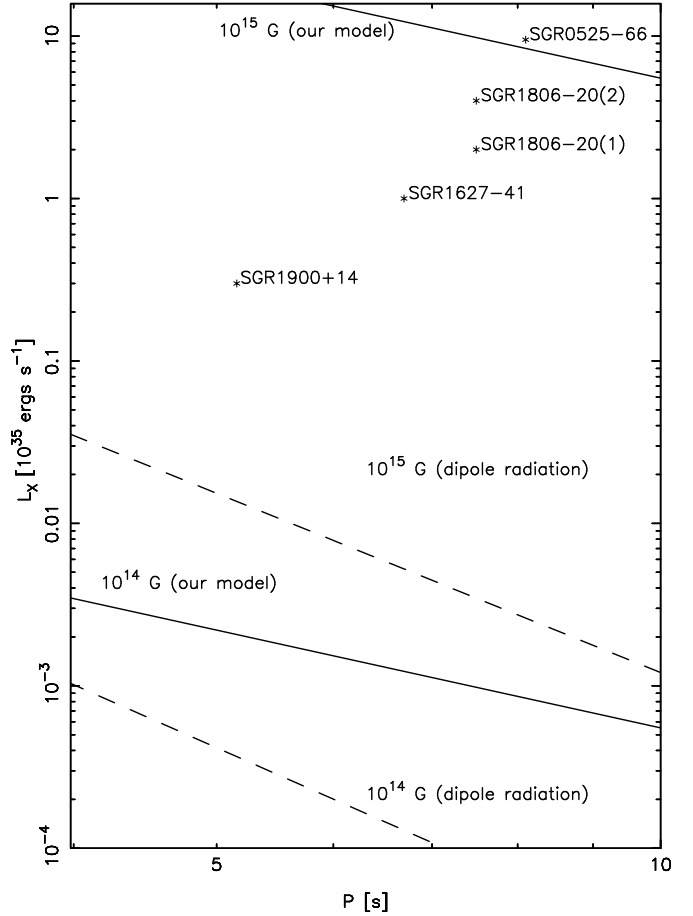}}
\caption[Lx-P relation]
 {The X-ray luminosity versus period for the four SGRs for which these values have been measured (from Kaplan 2000, and references
 therein). The solid lines are predictions given by equation~(\ref{eq:qluminosity}) for stars with magnetic fields of $10^{15}\, {\rm G}$ (upper curve) and $10^{14}\, {\rm G}$, for $R = 10\, {\rm km}$ and $M = 1M_{\sun}$. The dashed lines are predictions given by equation~(\ref{eq:qluminosity_standard}) for stars with magnetic fields of $10^{15}\, {\rm G}$ (upper curve) and $10^{14}\, {\rm G}$, for $R = 10\, {\rm km}$.}
\label{fig:LxP}
\end{figure}
\begin{eqnarray}
L_{\rm dip} & = & - \frac{4 B^2 R^6 \Omega^4}{9 c^3} \simeq 2.6 \times 10^{36} {\rm ergs}\, {\rm s}^{-1} \times \nonumber \\ 
& \times & \left(\frac{B}{10^{15} {\rm G}}\right)^2 \left(\frac{R}{10 {\rm km}}\right)^6 \left(\frac{1 {\rm s}}{P}\right)^4\ .
\label{eq:qluminosity_standard}
\end{eqnarray}
\noindent
We note that for $P > 1\,{\rm s}$ we can still account for the observed SGR quiescent luminosity 
 with $B \leq 10^{15}\, {\rm G}$ while much higher fields are required to power the luminosity with dipole radiation (Figure 4).  

In our model, the quiescent phase emission consists of two sources of energy, related to vortex annihilation and dipole radiation. 
It would be interesting to compute the corresponding spectra and compare them to the now well-established best fit model of SGR 
spectra which is a composite of blackbody and power-law emission (Kaplan 2002; Kulkarni et al.\ 2003).  This is beyond the scope 
of this paper and is left as an avenue for future work. 

\section{Later bursts: {\it Crustquakes and debris impacts}} 
\label{sec:laterbursts}

\subsection{Crust formation and crustquakes}

A crust of hadronic matter suspended above
the quark star surface (Alcock, Farhi, \& Olinto 1986) might eventually form, leading to possible starquakes. 
Since known SGRs are solitary objects, such a crust should have formed due to fall-back of supernova ejecta. 
Also in the Quark-Nova (Ouyed et al. 2002) picture,
where a hot quark star is formed, high accretion rates are expected from
the fallback material (Ker\"anen\&Ouyed 2003).
Later bursts could be driven by energy released 
during crust fractures that result from magnetic stresses as
described in Thompson\&Duncan (2001). In the case of the neutron star crust, the corresponding energy is 
\begin{eqnarray}\label{Eq3}
\delta E_{\rm mag} &\simeq& 10^{44}\,
\left({\psi^2\over (10^{-2})^2}\right)\,
\left({B\over 10^{15}\ {\rm G}}\right)^{-2}\times\nonumber\\
&\times&\left({\rho_{\rm crust}\over 0.6\,\rho_{\rm N}}\right)^{1.6}\,
\left({\Delta R_\mu\over 0.3~{\rm km}}\right)^2\hbox{ergs}\ ,
\end{eqnarray}
where $\psi$ is the shear strain, $\rho_{\rm N}$ the nuclear saturation density
and $\Delta R_\mu$ is the extent of the fracture in the crust. If a similar phenomenon can
be expected for quark star crusts, we expect later (crust induced) bursts 
with energies as high as $\sim 10^{40}$ ergs (the maximum density for a quark star 
crust is close to the neutron drip value, $\rho_{\rm crust}\sim 4\times 10^{11}$ g cm$^{-3}$ 
see e.g. Glendenning 1997).

\subsection{Hard emission bursts}

Rare ($<1$\%) subsequent bursts of hard emission (Woods et al.\ 1999) could be due to random debris impacts. Again,
as an example, such debris could naturally result during disk (and the subsequent
planetary) formation around the quark star following an expulsion of the crust (Ker\"anen\&Ouyed 2003).

\section{SGRs in our model} 
\label{sec:SGRdisc}

\subsection{SGR 0526-66}

Applying our model to the SGR responsible for the famous 1979 March 5 giant burst, and for $P = 8\, {\rm s}$, we find the 
luminosity in the quiescent phase to be $8.6 \times 10^{35}\, {\rm ergs}\, {\rm s}^{-1}$ for $B = 10^{15}\, {\rm G}$. The 
corresponding minimum age, making use of equation~(\ref{eq:spindown}), is $\sim 3 {\rm kyr}$. Hence, our model does not 
require extreme magnetic fields ($> 10^{15}\, {\rm G}$) and predicts ages of at least a few kyr, in accordance with age 
estimates of the nearby SNR N49 (e.g., Vancura et al.\ 1992). However, we note that the association with SNR N49 is tenuous, and 
the large offset between the two objects would still be a challenge if the CFL phase transition occurred as recently as 1979 (see \S \ref{sec:formation}).  
Following its long quiescent phases SGR 0526-66 could
have acquired a thin  crust (\S \ref{sec:laterbursts}) which could
explain why it currently shows X-ray characteristics similar to AXPs (Kulkarni et al.\ 2003), two decades after becoming quiescent. We speculate that the 1979 March 5 giant burst, detailed in Table~\ref{tab:burstprop}, 
 could have been a signature of a CFL phase transition and associated physical processes, as described in 
\S \ref{sec:expulsion}. This SGR, displaying a giant burst and a subsequent active phase followed by a long 
quiescence, fits best within the picture outlined in this paper. 

\subsection{SGR 1900+14}

Because of the presence of SGR burst activity prior to the main event 
(Hurley et al.\ 1999), 
the 1998 August 27 event in SGR 1900+14 would be best explained by, 
 e.g., restructuring of the magnetic field (\S~\ref{sec:vortices}), crustquakes or debris impact 
(\S~\ref{sec:laterbursts}).

\section{Discussion}
\label{sec:discussion}
\subsection{The suggested SGR-AXP connection}

It has been debated whether anomalous X-ray pulsars (AXPs) and SGRs are 
connected, with both classes of systems being magnetars 
(e.g., Chatterjee, Hernquist \& Narayan 2000; 
Gavriil, Kaspi, \& Woods 2002). AXPs display persistent strong 
X-ray emission with $P \sim 6-12$~s
pulsations. The term ``anomalous'' comes from the fact that the X-ray emission is not powered by
rotational energy or by accretion from a companion star (AXPs are solitary objects).
The likely association of three AXPs with supernova remnants indicate that
they are young ($t < 10^4$~yr) systems with unusually fast spin-down rates 
(Gaensler et al.\ 2001).
Like SGRs, the AXPs are rare objects (five confirmed cases are currently known), 
and they have a similardistribution of rotational periods.

If AXPs are magnetars, their emission is most likely powered by the decaying
magnetic field. Recent observational results have suggested a link between
the two classes of objects: Gavriil et al.\ (2002) report SGR-like X-ray bursts
from the AXPs 1E 1048.1-5937 and 1E 2259.1+586, and Kaspi et al.\ (2003) report a major 
SGR-like X-ray outburst from 1E 2259.1+586. Furthermore, Kulkarni et al. (2003) report
{\it Chandra} observations of SGR 0526-66 in a quiescent phase showing that the object 
has X-ray properties similar to an AXP. These similarities have been suggested to 
favour a common magnetar model for AXPs and SGRs (see Kaspi 2004
for a recent discussion). However, it remains to be seen if these 
two classes of objects are similar physical systems that could perhaps be linked 
through a simple evolutionary model, or whether they are in fact disparate.

In our model, SGRs are strange matter quark stars that have
undergone the phase transition into a colour superconducting state. Based on this, we
cannot identify a simple connection between AXPs and SGRs, but our model merely 
suggests that they are not closely connected. We discuss below how missing glitch activity 
and inferred transverse velocities of SGRs could hint in favour of our picture.

\subsection{Glitches in our model}

Glitch activity is apparent in AXPs, but seems to be absent in 
SGRs, see Woods et al. (2003a and b) and Kaspi et al. (2003).   
Vortices are a natural outcome of the CFL phase transition, 
and the interaction between the vortices and the crust could in principle 
lead to glitch activity (see Alpar 1991). If only a tiny crust is expected 
for a CFL star, glitches would be unlikely unless a  
glitch mechanism which does not involve the crust-core interaction 
is involved. A possibility for such a mechanism has been suggested by 
Alford, Bowers and Rajagopal (2001).  They propose that in a certain 
range of densities quarks may form Cooper pairs with nonzero momentum, 
leading to pairing energy gaps which vary periodically in a crystalline 
pattern, and this structure opens up the possibility of vortex pinning.  
However, the details of this mechanism have not yet been worked out.   
Thus we do not expect glitch activity to occur in SGRs in our model
except maybe in a case where a sizeable crust has been formed. This
might be possible e.g. if the hadron-quark phase transition has happened 
very quickly after the supernova explosion and the crust has been 
formed from the possible fall-back matter of the supernova ejecta.

\subsection{Transverse velocities}

At least half of the AXPs are now confirmed to be located near the centers of  
supernova remnants (SNRs) (see
discussion by Gaensler 2002). This infers transverse velocities of 
$\le 500\, {\rm km}\, {\rm s}^{-1}$ and an age of $\le 10\, {\rm kyr}$. 
Furthermore, three out of the five confirmed SGRs are located near SNRs, 
but the SGR separation 
from the SNR centers and the SNR ages imply transverse velocities of order 
$\geq 1000\, {\rm km}\, {\rm s}^{-1}$, which is higher than what would be expected 
from a SN explosion alone. In our model, an neutron star undergoes a phase transition into 
quark matter. This phase transition is accompanied by a neutrino emission (see e.g.~Iwamoto 1980). If the neutrino emission is not 
fully isotropic, this may lead into a high transverse velocity of the  quark star,
explaining the large offsets between some SGRs and apparently associated 
nearby young SNRs.

\subsection{Formation scenarios}
\label{sec:formation}

The transition from
hadronic matter into quark matter can in principle happen immediately
during or after the supernova explosion,
 but also very much later than 
that. If formed quickly after the supernova occurred, the emitted gamma rays
would be absorbed by the expanding supernova ejecta, and 
there would be no detectable gamma-ray signature of the transition
from neutron star matter into quark matter. In this case, the resulting 
quark star would be born in a relatively dense evironment and might form 
a crust from accreted material. We would then expect subsequent 
SGR-type bursting behaviour to be caused by mechanisms discussed in 
\S~6. As discussed above, because of the crust, the star could show 
glitch activity and could in this respect have 
characteristics more similar to AXPs than other SGRs. 

The hadron-quark transition may also happen much later than
the supernova explosion. Quark star formation 
can, for example, happen as a result of accretion or spin-down of a 
neutron star.  In this case, the CFL 
phase transition should be accompanied by an observable gamma ray emission
(as a giant burst in our picture), as we speculate is the case for SGR0526-66. 
In this second scenario, the quark star would remain bare or have only a thin
crust due to low accretion (since the objects are solitary). Therefore there would
be no glitch activity. However, if indeeed SGR0526-66 is a CFL star, the late
formation and a resulting recent boost of the transverse velocity would not explain its current 
offset from the nearby young SNR N49: Even if the neutrino emission had 
been anisotropic, the SGR can not have moved very far during a few decades, and so the late 
formation scenario would suggest that there is no physical association between SNR N49 
and SGR0526-66. In any case, there is currently no compelling observational evidence for 
such an association (Gaensler 2002).  

It is important to notice that the CFL phase transition will occur only once, 
so the model discussed in this paper predicts that there will be only one 
giant burst.  Also, note that the probability of observing such an event 
is small, since the mean time between supernovae is about 100 years  
and formation of a quark star is maybe even rarer.  This should be 
compared to the short time we have had gamma-ray observatories.

It has been demonstrated that CFL stars with very short rotation periods should not exist, 
because they are prone to so-called r-mode instabilities (Madsen 2000). 
Since the CFL star we consider has a rotation period of the order of 
a few seconds, this should not be a problem for our model.

Finally, the era of decaying vortices can be 
estimated to last $10^3$ to $10^5$ years in our model. Even if the formation of the
quark star happens immediately after (or during) the supernova explosion, the 
SGR should remain active long enough to account for the observed $\sim 10\;{\rm s}$ periods.
A fast spin down occurs if the magnetic field is initially large as our model assumes.

\section{Conclusion}
\label{sec:conclusion}

We presented an alternative model for explaining SGRs where a newly
born quark star experiences bursting activity as it
 cools below the critical temperature for the onset of CFL.   
In our model, the subsequent magnetic field expulsion and reorganization
 are crucial ingredients in explaining the giant bursts
and the subsequent weaker ones; magnetic field strength and period alone cannot be 
responsible for the unusual properties of SGRs according to our model. 
We also discuss the relation between AXPs and SGRs which in our model
are separate objects.  

\begin{acknowledgements}
The research of R.O. is
supported by grants from the Natural Science and
Engineering Research Council of Canada (NSERC).
H.D. is funded by a post-doctoral fellowship 
from the  Research Council of Norway. 
H.D., {\O}.E. and R.O. acknowledge the hospitality of Nordita and
P.K. and R.O. the hospitality of the University of Oslo.
\end{acknowledgements}

\end{document}